# Ultra-High Thermoelectric Power Factors in Narrow Gap Materials with Asymmetric Bands

*Patrizio Graziosi,*[*,†,‡]  *Neophytos Neophytou*[†]

[†] School of Engineering, University of Warwick, Coventry, CV4 7AL, UK

[‡] Consiglio Nazionale delle Ricerche – Istituto per lo Studio dei Materiali Nanostrutturati, CNR – ISMN, via Gobetti 101, 40129, Bologna, Italy

ABSTRACT. We theoretically unveil the unconventional possibility to achieve extremely high thermoelectric power factors in lightly doped narrow gap semiconductors with asymmetric conduction/valence bands operated in the bipolar transport regime. Specifically, using Boltzmann transport simulations, we show that narrow bandgap materials, rather than suffering from performance degradation due to bipolar conduction, if they possess highly asymmetric conduction and valence bands in terms of either effective masses, density of states, or phonon scattering rates, then they can deliver very high power factors. We show that this is reached because, under these conditions, electronic transport becomes phonon scattering limited, rather than ionized impurity scattering limited, which allows large conductivities. We explain why this effect has not been observed so far in the known narrow-gap semiconductors, interpret some recent related experimental findings, and propose a few examples from the half-Heusler materials family for which this effect can be observed and power factors even up to 50 mW/mK$^2$ can be reached.

## 1. Introduction

Thermoelectric materials convert heat from temperature gradients directly into electricity and they can contribute to energy sustainability and reduction in the use of fossil fuels. [1] Despite their enormous possibilities, their large-scale applicability is hindered by the high costs and high toxicity of the highest performance materials. Large efforts are undergoing, however, to discover new, more efficient and/or non-toxic thermoelectric (TE) materials, both from the experimental and computational sides. [1-3]

The dimensionless figure of merit $ZT = \frac{\sigma S^2}{k_L + k_e} T$ quantifies the TE material conversion efficiency, where $\sigma$ is the electrical conductivity, $S$ the Seebeck coefficient, $k_L$ and $k_e$ are the lattice and electronic thermal conductivities, respectively, and $T$ is the absolute temperature. Materials research to date revolved around nanostructuring to reduce the thermal conductivity, [4-8] but several strategies to improve the power factor PF = $\sigma S^2$ [1, 9-14] and complex bandstructure materials [3, 15, 16] are attracting significant interest as well. As a matter of fact, an analysis of the experimental literature on more than 11,000 compounds, [9, 17] reveals that the materials with the highest $ZT$ are generally found more often amongst the materials with high power factors. Thus, the detailed material features that lead to high PF values are a valuable resource, [18] especially if combined with so-called "inverse design" strategy, where a target for the desired property of a new material is set, and theoretical calculations are then used to identify the compounds that exhibit that property. [19, 20]

Narrow bandgap can provide increased carrier density due to thermal excitation form the valence band (VB) to the conduction band (CB) compared to large gap materials, which increases the number of majority carriers in the CB and add minority carriers in the VB. This improves the electrical conductivity, but also leads to a decrease of the Seebeck coefficient that cancels out any benefit. As a result, a bandgap $E_G$ of at least ~ 10 $k_B$T is usually suggested to achieve good TE performance. [21] In this work, we show that in narrow gap materials, despite bipolar effects usually leading to performance degradation, it is possible to achieve extremely high power factors. This happens under two conditions: i) lightly doped conditions with the Fermi level residing within the bandgap, and ii) when the conduction and valence bands of the material are highly asymmetric, in terms of various quantities: effective masses, density of states, scattering rates, etc.

For the former condition, under lightly doped conditions, the Fermi level is pushed away from the bands into the bandgap of the material and the electron mobility approaches the intrinsic phonon-limited value, which is higher compared to the dopant ionized impurity scattering limited value (as is the usual case in the highly doped TE materials at optimal conditions). The presence of the minority band in narrow bandgap semiconductors, however, does not allow the Fermi level to shift away from the bands beyond the intrinsic level as the dopant density is reduced, which leads to nearly intrinsic conditions with high majority and sizeable minority carrier concentrations, hence high conductivity and phonon-limited mobility. The second condition dictates the asymmetry of the bands. Asymmetric CB and VB make the bipolar zero crossing point of the Seebeck coefficient to be non-coincidental with the intrinsic level of high conductivity. Thus, high conductivity is achieved at finite Seebeck coefficient, which leads to a large power factor.

The paper is structured as follows: In Section 2 we present the computational methods and underline the differences between unipolar and bipolar calculations, especially with respect to the position of the Fermi level at a given doping density. In Section 3, we present the conditions under which the large PFs are observed. In Section 4 we discuss the advantages of this large PF resulting from the reduced doping, explain why this effect has not been observed so far in the known narrow-gap thermoelectric materials, and propose materials from the half-Heusler group in which this effect can be detected. Finally, in Section 5, we conclude by summarizing our findings and highlighting their impact on the TE community.

2. **Theoretical methods**

For computing the thermoelectric coefficients, we use the Boltzmann Transport Equation (BTE) in the relaxation time approximation, considering the full energy dependence of the relaxation times. [13, 18, 22-31] We consider first analytical parabolic bands to explore and optimize the novel effect we investigate, and then a full-band numerical approach to compute the TE coefficients in half-Heusler materials. [18, 23]

In the BTE formalism, the electrical conductivity, $\sigma$, and the Seebeck coefficient, $S$, tensors, are expressed as

$$\sigma_{ij} = q_0^2 \int_E \Xi_{ij}(E) \left(-\frac{\partial f_0}{\partial E}\right) dE \tag{1a}$$

$$S_{ij} = \frac{ek_B}{\sigma} \int_E \Xi_{ij}(E) \left(-\frac{\partial f_0}{\partial E}\right) \frac{E-E_F}{k_B T} dE \tag{1b}$$

where $E_F$, $T$, $e$ $k_B$, are the Fermi level, the absolute temperature, the electronic charge, and the Boltzmann constant, respectively, $f_0$ is the equilibrium Fermi distribution, and $i, j$ are the Cartesian directions (in all the calculations we use, $i = j = x$). $\Xi(E)$ is the transport distribution function (TDF) defined as [25, 26, 32, 33]

$$\Xi_{ij}(E) = \sum_{k,n}^{BZ} v_{i(k,n)} v_{j(k,n)} \tau_{i(k,n)} \delta_{(E_{k,n}-E)} \tag{2}$$

In equation (2) $v_{i(k,n,E)}$ is the $i$-component of the band velocity of the state that contributes to transport defined by the wave vector $\mathbf{k}$ in the band $n$ at energy $E$, $\tau_{i(k,n,E)}$ is the anisotropic momentum relaxation time [22, 28-30, 34] (which combines the relaxation times of each scattering mechanism, defined below, using Matthiessen's rule). The TDF, $\Xi(E)$, is defined over the whole energy interval of interest, encompassing both the valence and conduction bands so that a unique integral over the energy interval is performed. [18, 35]

### 2.1 *Analytical band simulations*

For isotropic and parabolic bands, the density of states and the band velocity can be expressed analytically as a function of energy, thus the TDF for the transport direction *i* becomes

$$\Xi_i(E) = v_{i(E)}^2 \tau_{i(E)} g_{(E)} \tag{3}$$

where $1/\tau_{i(E)} = \sum_{m_s} 1/\tau_{i(E)}(m_s)$ is the overall relaxation time, with the sum running over all the scattering mechanisms $m_s$, that we consider. $v_{(E)}$ is the carrier velocity and $g_{(E)}$ the density of states (DOS).

We consider here only elastic scattering without loss in genericity in the conclusions. Elastic scattering with acoustic phonons, in the acoustic deformation potential theory (ADP), is used as representative of all the electron-phonon processes, [35] in order to reduce the number of simulation parameters. In addition, we consider the scattering with ionized impurities (IIS) and eventually a simplified contribution of scattering due to the grain boundaries (GB). Thus, the relaxation time for the scattering mechanisms considered are: [36, 37]

$$\frac{1}{\tau_{(E)}^{(ADP)}} = \frac{\pi D^2 k_B T}{\hbar \rho u_s^2} \frac{1}{2} g_{(E)} \tag{4a}$$

$$\frac{1}{\tau_{(E)}^{(IIS)}} = \frac{e^4 N_{imp}}{16\pi \varepsilon^2 \sqrt{2m_{DOS}} E^{3/2}} \left( \ln(1+\gamma^2) - \frac{\gamma^2}{1+\gamma^2} \right) \tag{4b}$$

$$\frac{1}{\tau_{(E)}^{(GB)}} = v_{(E)}/L_G \tag{4c}.$$

In equation (4a) $D$ is the deformation potential for acoustic phonons, $\rho$ is the mass density, and $u_s$ is the speed of sound. In equation (4b) $e$ is the electron charge, $N_{imp}$ is the dopant impurity density, $\varepsilon$ is the dielectric constant ($\varepsilon = \varepsilon_r \varepsilon_0$ with $\varepsilon_0$ the vacuum permittivity and $\varepsilon_r$ the relative dielectric constant), $m_{DOS}$ is the DOS effective mass, and $\gamma^2 = 8m_{DOS}EL_D^2/\hbar^2$ is the screening term that contains the screening length $L_D = \sqrt{\frac{\varepsilon_r \varepsilon_0}{e}\left(\frac{\partial n}{\partial E_F}\right)^{-1}}$ where $E_F$ is the Fermi level and $n$ is the carrier density. [23] In equation (4c) $L_G$ is the average grain size considered and $v_{(E)}$ the carrier velocity, constituting the simplest treatment of boundary scattering.

We consider here isotropic 3D materials, for which the DOS is $g_{(E)} = \frac{m_{DOS}^{3/2}}{2\pi\hbar^3}\sqrt{2E}$ and the carrier velocity is $v_{(E)} = \sqrt{2E/m_c}$, where $m_c$ is the conductivity effective mass. As transport is along one crystallographic direction, following standard approach, [37, 38] we consider that $\left|v_{(E)}\right|^2 = \sum_i v_{i(E)}^2$ where $i$ runs over the Cartesian coordinates, such that within the TDF in equation (3) we use $v_{i(E)}^2 = \frac{1}{3}2E/m_c$. We compose the relaxation times in equation (4) according to the Matthiessen's rule and obtain an overall $\tau(E)$ to be used in the TDF. This analytical approach is used to extract the data plotted in Figures 1 to 5.

*2.2 Full-band approach*

Complex bandstructure TE materials have highly non-parabolic warped bands, which requires a fully numerical treatment in the computation of their TE coefficients. [18, 23] The TDF then becomes

$$\Xi_{ij}(E) = \frac{2}{(2\pi)^3} \oiint_{\mathfrak{L}_E^n} v_{i(\mathbf{k},n,E)} v_{j(\mathbf{k},n,E)} \tau_{i(\mathbf{k},n,E)} \frac{1}{\hbar} \frac{dA_{\mathbf{k},n,E}}{|v_{(\mathbf{k},n,E)}|} \qquad (5)$$

where the surface integral is over the constant energy surface $\mathfrak{L}_E^n$ for the band of index *n*. In equation (5) $v_{i(k,n,E)}$ is the *i*-component of the band velocity of the transport state defined by the wave vector **k** in the band *n* at energy *E*, and $\tau_{i(k,n,E)}$ is its combined momentum relaxation time. $dA_{k,n,E}$ is the surface area element associated with the (**k**,*n*,*E*) transport state, $v_{(k,n,E)}$ is its band velocity and $\frac{1}{\hbar}\frac{dA_{k,n,E}}{|v_{(k,n,E)}|}$ its DOS.[23] For each transport state and each scattering mechanism $m_s$, the relaxation time $\tau_{i(k,n,E)}^{(m_s)}$ is given by a surface integral on the final constant energy surface $\mathfrak{L}_{E'}^{n'}$ as

$$\frac{1}{\tau_{i(k)}^{(m_s)}} = \frac{1}{(2\pi)^3} \oiint_{\mathfrak{L}_{E'}^{n'}} |S_{k,k'}^{(m_s)}| \frac{dA_{k'}}{\hbar |v_{(k')}|} \left(1 - \frac{v_{i(k')}}{v_{i(k)}}\right) \quad (6)$$

where **k** and **k'** represent the initial and possible final states. The symbols **k** and **k'** in equation (6) include the band indexes *n* and *n'* and state energy *E* and *E'*, respectively. Here when $E' = E$ the scattering is elastic and when $E' = E \pm \hbar\omega$ the scattering is inelastic involving a phonon of frequency $\omega$. $|S_{k,k'}|$ is the transition rate between **k** and **k'**, and $v_i$ is the carrier velocity. The term $\left(1 - \frac{v_{i(k')}}{v_{i(k,n,E)}}\right)$ is an approximation for the momentum relaxation time.[13, 22, 23, 25, 26, 28, 34] $|S_{k,k'}|$ is derived from Fermi's Golden Rule for the different scattering mechanisms, and the relevant scattering parameters are taken from first-principle calculations. As in the analytical band case, we employ the deformation potential theory for electron-phonon scattering (acoustic and optical), and ionized impurity scattering with charge screening. For the longitudinal optical phonon frequency, we adopt an averaged value that is assumed to be constant over the reciprocal lattice unit cell. The numerical treatment is here applied to half-Heusler compounds, using scattering parameters from literature [39-42] and both elastic and inelastic processes for electron-phonon scattering.[23] As in the analytical treatment case, the TDF is defined over the whole energy range of valence and conduction bands and then integrated over. More details on this full-band numerical scheme is described in detail in Ref. 23 and is used to obtain the data in Figure 6.

2.3 *Unipolar versus bipolar transport*

Under unipolar transport considerations there is only one type of carrier, for example electrons in the CB, and their density is given by $n = \int g_{(E)} f_{0(E,E_F,T)} dE$ where $f_0$ is the Fermi Dirac distribution and the integral extends over the CB energy range. The density of ionized impurities is considered to be the same as the density of mobile carriers, i.e. $N_{imp} = n$. Under bipolar considerations, for the same Fermi level position, the electron density $n$ is as defined above, but a hole density $p$ is also present. Because charge neutrality dictates that the total ionized impurity density is equal to the total density of mobile charges, $N_{imp} = p - n$, the presence of minority carriers (holes in this case) will force the number of dopant impurities, at a specific Fermi level position, to always be lower compared to what it would be in the unipolar case. Thus, a large difference in $N_{imp}$ can exist between unipolar and bipolar considerations for the same majority band and Fermi level position, due to the presence of the minority band. In particular, under bipolar considerations, an intrinsic Fermi level exists, so that $n = p$ and hence $N_{imp} = 0$, a situation that is never encountered in unipolar material cases.

The usual bipolar considerations encountered in the literature when calculating the combined electrical conductivity and Seebeck coefficient, treat the CB and VB separately as being unipolar materials with infinite bandgap, obtaining the electrical conductivities and Seebeck coefficients $\sigma_e$, $S_e$, $\sigma_h$, $S_h$ for electrons in the CB and holes in the VB, respectively. Afterwards, these are combined to obtain the bipolar transport coefficients $\sigma_b$, $S_b$ as: [43]

$$\sigma_b = \sigma_e + \sigma_h \tag{7a}$$

$$S_b = \frac{\sigma_e S_e + \sigma_h S_h}{\sigma_e + \sigma_h} \qquad (7b).$$

Nevertheless, this cannot always be correct, because the impurity density $N_{imp}$ (at a certain Fermi level) differs between unipolar and bipolar considerations (in the unipolar case, $N_{imp} = n$, whereas in the bipolar case, $N_{imp} = n - p$). Thus, unipolar and bipolar considerations may lead to different $N_{imp}$ and hence different ionized impurity scattering rate and different conductivity. This is particularly important in narrow-gap materials.

3. **Results and discussion**

In this section we describe the conditions for achieving high power factors in narrow bandgap materials using analytical parabolic bands. Afterwards, we explore this possibility for realistic half-Heusler materials using numerical DFT extracted bandstructures.

3.1 *Ultra-high power factor in narrow bandgap materials*

Our analysis compares analytical parabolic bands for symmetric "S" and asymmetric "A" bipolar materials. Their TE transport coefficients are presented in Figure 1a-c and the bandstructures are depicted in Figure 1d. The "S" bandstructure consists of a parabolic CB and VB of effective masses equal to the rest electron mass $m_0$, while in bandstructure "A" the effective mass of the CB is 0.5 $m_0$ and the mass of the VB is kept at $m_0$. The bandgap is considered to be 0.2 eV and the temperature 1000 K, as the effect we investigate is pronounced when bipolar effects are strong – we examine lower temperatures further below. The electrical conductivity, $\sigma$, is shown in Figure 1a, the Seebeck coefficient, $S$, in Figure 1b, and the power factor, PF = $\sigma S^2$, in Figure 1c. We plot the TE coefficients versus the relative position of the Fermi level, $\eta_F$, which has an one-to-one correspondence with the doping density. $\eta_F$ is set to 0 in the middle of the bandgap. The calculations are performed assuming a complete bipolar transport scheme with energy integration over the whole energy range, as explained in section 2, and we consider both ADP and IIS scattering mechanisms. The deformation potential value is set to 2 eV and the relative dielectric constant to 20 (typical values for half-Heusler materials, for example). [18, 23, 42]

The red dash-dot lines are simulation data performed for the phonon-limited transport case in the "A" bandstructure, (considering only ADP). The phonon-limited results (red-dashed lines) in Figure 1a-c show the usual behavior. In Figure 1a the conductivity increases as the Fermi level is pushed into the valence band to the left, or the conduction band to the right. When the Fermi level is in the bandgap the conductivity is at its minimum, whereas the line is not symmetric as the bandstructure is not symmetric. The Seebeck coefficient in Figure 1b also has the usual behavior with a zero crossing slightly closer to the valence band because Seebeck crosses zero when the polarity is inverted. This happens when $\sigma_e = \sigma_h$ (equivalently when $n\mu_e = p\mu_h$, where $\mu_{e/h}$ is the electron/hole mobility). [21] As the effective mass of holes is larger in the "A" structure, which makes its conductivity (and mobility) lower, to satisfy the polarity reversal condition $\sigma_e = \sigma_h$, the Fermi level needs to be closer to the VB than to the CB.

In the case where both phonon and ionized impurity scattering is considered (black and blue lines for "S" and "A", respectively), a very different behavior is observed. The electrical conductivity, $\sigma$, surprisingly spikes in the vicinity of the intrinsic Fermi level position, $E_F^{intr.}$, where the carrier density of electrons, $n$, and holes, $p$, are the same, i.e. $n = p$. As the impurity density is reduced for Fermi levels approaching $E_F^{intr.}$ towards the midgap, ionized impurity scattering weakens significantly, electron-phonon scattering becomes dominant, and $\sigma$ tends to reach the phonon-limited value when the Fermi level is close to $E_F^{intr.}$. This is more evident in Figure 1a, where the conductivity of the "A" structure with phonon plus IIS (blue line), reaches that of the phonon-limited material (red-dashed line). A similar (but lower) peak is observed for the symmetric band "S" material, but in this case $E_F^{intr.}$ is at the midgap due to the band symmetry (black line). The $N_{imp}$ versus Fermi level position is shown in Figure 1e for the "S" bandstructure (black line). Clearly, for the Fermi level to be exactly at the midgap, the material needs to be undoped and $N_{imp}$ is zero –indicated by the drop to very low values in Figure 1e. Notice that this effect appears only in bipolar materials with narrow bandgaps, where the presence of the minority carries allows the same number of majority carries, but with reduced $N_{imp}$, and the presence of the minority band does not allow the Fermi level to shift away from the majority band with reduced $N_{imp}$. The dashed lines in Figure 1e show the corresponding doping density in a unipolar material, which it is always higher in the midgap region. In a unipolar material, the Fermi level shifts away from the bands monotonically with reduced $N_{imp}$, such that by the time phonon-limited transport conditions are reached, the conductivity and PF are significantly reduced.

Thus, the spike in the electrical conductivity appears when $n = p$, which happens when the Fermi level is at the intrinsic position, $E_F^{intr.}$. In the case of the "S" bandstructure, because of the complete symmetry of the conduction and valence bands (CB and VB), the intrinsic Fermi level position is at midgap, $E_F^{intr.} = 0$. In the case of the "A" bandstructure, the different effective masses in the CB and VB make the density of states (DOS) also asymmetric. Inevitably $E_F^{intr.}$ is not in the middle of the gap, but it is located closer to the CB which has the smaller DOS (smaller effective mass). Note that the conductivity is not only benefitted by approaching the phonon-limited value, but compared to unipolar materials it is even higher because it also has the contribution of the minority carriers.

The location of $E_F^{intr.}$, and its shift towards the low DOS band in asymmetric bandstructures is key in translating this high, phonon-limited conductivity to high PFs. In symmetric CB/VB bandstructure materials $E_F^{intr.}$ is at midgap, at the same point where the Seebeck coefficient changes polarity and experiences a zero crossing (black line in Figure 1b). Thus, at the point where the conductivity spikes (black line in Figure 1a), the PF reduces to zero (black line in Figure 1c). Importantly, this is not the case for the asymmetric CB/VB material case. The conductivity spike appears at $E_F^{intr.}$ when $n = p$ and $N_{imp} = 0$, but the Seebeck coefficient changes polarity when $n\mu_e = p\mu_h$, where $\mu_{e/h}$ is the electron/hole mobility. These two conditions occur at different Fermi level positions in asymmetric bandstructures, and thus, the finite Seebeck coefficient at $E_F^{intr.}$ where the conductivity is large, allows for large PFs as well, reaching the phonon-limited PFs (blue and red-dashed lines in Figure 1c). Note that the point where the Seebeck coefficient crosses the zero in the "A" bandstructure is different under phonon-limited and phonon + IIS conditions, because of the different role that different masses play on the relaxation times. Thus, for the specific example of the "A" material we consider in Figure 1, where the DOS in the CB is smaller, the peak of $\sigma$ at $\eta_F = E_F^{intr.}$ appears closer to the conduction band. As a result, of the conductivity peaking at finite Seebeck, the PF as well peaks around $\eta_F = E_F^{intr.}$, i.e. blue line in Figure 1c.

Our calculations in Figure 1a-c and 1e are performed at $T = 1000$ K. It is interesting, however, to observe the behavior of the intrinsic Fermi level $E_F^{intr.}$ with temperature for the "A" and "S" bandstructures, as this is what determines the peak of the PF. This is shown in Figure 1f. With CBM and VBM we denote the Conduction Band Minimum and Valence Band Maximum, respectively. The bandgap of 0.2 eV is considered to be constant with temperature. For the "S" bandstructure, $E_F^{intr.}$ is shown by the black-diamond solid line, while for the "A" bandstructure is shown by the blue-stars solid line. As expected, in the symmetric material the $E_F^{intr.}$ is always in the midgap, whereas for the asymmetric CB/VB material, it is closer to the lighter band and moves even closer with increasing temperature. Thus, it will be expected that the conductivity spike in Figure 1a will shift to the right, closer to the light mass CB with temperature as well.

The fixed $E_F^{intr.}$ in the symmetric material and the upward shift of the $E_F^{intr.}$ of the asymmetric structure with temperature is a unique feature of the bipolar material and a consequence of charge neutrality. For a unipolar material the $n = p$ condition cannot be satisfied and the Fermi level position shifts away from the corresponding band (CB or VB) with temperature in order to keep charge neutrality by keeping the number of mobile carriers equal to $N_{imp}$. Essentially the broadening of the Fermi distribution with temperature is compensated with a shift of the Fermi level away from the bands. The unipolar $E_F$ position is shown in Figure 1e with dotted lines for a carrier density of $10^{18}$ cm$^{-3}$, (dotted blue for the CB and dotted red for the VB). Because of the absence of a minority band, the Fermi level shifts arbitrarily far from the CBM/VBM upon increasing temperature, [23] whereas bipolar effects stabilize the Fermi level position. [18, 35]

At this point it is interesting the compare the PF using the usual combination of the unipolar coefficients of the CB and VB as in equation (7), versus the full bipolar evaluation of the TE coefficients in the presence of IIS. The failure of the combined unipolar transport model to account for the correct impurity density, does not adequately describe the charge transport coefficients when the Fermi level is in the bandgap, the region where we are interested in. In Figure 2 we compare the TE coefficients computed using full bipolar considerations with the ones computed by considering the two unipolar cases (for the CB and VB) combined. Figures 2a, 2b and 2c (left column) are for the "S" bandstructure, where the results from the bipolar calculations are shown in black lines. Figures 2d, 2e and 2f (right column) are for the "A" bandstructure, with the results from bipolar calculations shown in blue lines. The green dotted lines are for results obtained using separate unipolar calculations for each band, which are then combined to provide the red dotted lines. Figure 2 clearly shows that the combined unipolar calculations (red dotted lines) can capture the transport coefficients away from the bandgap where the bipolar effects are negligible. These are the heavily degenerate conditions with the $\eta_F$ deep into the bands. They fail to describe the transport coefficients when the bipolar effects are important. Of course, if one computes the unipolar coefficients using the correct $N_{imp}$ (extracted from consideration of the minority band) then the combined unipolar results become identical to the bipolar results, but otherwise, the trends and amplitudes for the TE coefficients are very different.

It is noticeable, especially for the asymmetric bandstructure "A", which additionally exhibits a PF spike, that the PF in the bipolar case (black lines) is larger compared to the PF of the combined unipolar cases (red dotted lines) near the midgap level. Achieving large PFs for Fermi level positions near the intrinsic Fermi level, can be advantageous as doping is often a challenging step in optimizing TE materials, and in the asymmetric CB/VB case, intrinsic materials seem to perform as good, or even better, compared to highly doped materials. Thus, the difference between the blue and black lines earlier in Figure 1c is what benefits asymmetric CB and VBs provide to the PF, and the difference between the blue line and the red line in Figure 2f is the degree of the improvement that could be neglected if full bipolar treatment and accurate $N_{imp}$ is not considered in the calculation.

### 3.2 *Criteria and parameters for high PFs*

From here on, we investigate influence of the bandstructure and transport parameters that can contribute to creating an asymmetry in the bands, and how that affects the PF. We begin with exploring the specifics of the anisotropy in the bandstructure, which can be created by allowing anisotropy in the density of states (DOS) effective mass, $m_{DOS}$, or the conductivity effective mass, $m_\text{c}$. We leave the VB masses at $m_\text{DOS} = m_\text{c} = 1$ and change the CB masses. Anisotropic valleys have different values for the effective masses $m_x$, $m_y$, $m_z$, which results in different values for the DOS effective mass, $m_\text{DOS} = \sqrt[3]{m_x m_y m_z}$, and for the conductivity effective mass, $m_\text{c} = \frac{3}{1/m_x + 1/m_y + 1/m_z}$. [37] In Figure 3 we compare the TE coefficients at $T = 1000$ K for three material cases in which the masses and bands in the CB are as shown in Figure 1d: i) when the CB masses are $m_\text{DOS} = 0.5$ and $m_\text{c} = 0.5$ (blue lines), ii) when the asymmetry is limited to the DOS effective mass, i.e. $m_\text{DOS} = 0.5$ and $m_\text{c} = 1$ (red lines, labelled "A$_\text{D}$"), and iii) when the asymmetry is limited to the conductivity effective mass, i.e. $m_\text{c} = 0.5$ and $m_\text{DOS} = 1$ (green lines labelled "A$_\text{C}$"). These bandstructures are also depicted in Figure 1d and labelled accordingly.

Clearly, from Figure 3a we observe that it is the asymmetry in $m_\text{DOS}$ which places the spike further from the midgap (blue and red lines with $m_\text{c} = 0.5$), reflecting the role of the DOS. The different asymmetries, however, also have an influence on the zero crossing of the Seebeck coefficient in Figure 3b, for which the necessary condition is $n\mu_e = p\mu_h$. This depends on both $m_\text{DOS}$, which affects $n$ and $p$, as well as $m_\text{c}$, which affects the mobilities. [18] Thus, the three bandstructures feature different values for $S$ at the intrinsic Fermi level positions (at the spikes in Figure 3a). Consequently, the PF in Figure 3c is affected differently. A small spike is detected in the PF when the asymmetry is in the $m_\text{DOS}$ (red line). However, since the PF amplitude depends on $m_\text{c}$ rather than on $m_\text{DOS}$, [18] and $m_\text{c}$ in this case is large, the PF is also lower. This is reinforced by the fact that the smaller CB $m_\text{DOS}$ forces the zero crossing of the Seebeck coefficient closer to the conductivity spike. On the contrary, when the asymmetry is in the $m_\text{c}$, and $m_\text{c}$ is lower, the overall PF is higher, but the spike in the low doping region near the midgap is absent (green line). Although the PF in this case can be larger compared to the magnitude of the ''A'' structure at the spike, we show in Figures 4 and 5 that when the asymmetry increases further with the tuning of other transport parameters as well, the spike can reach very large PF values. Thus, to observe a high spike in the PF for $E_F$ around the intrinsic level, we need asymmetry in the $m_{DOS}$ and $m_c$ with lower mass values in the majority band.

Note that the way we use $m_{DOS}$ in the calculations has a direct influence on the DOS in a similar way that band degeneracy would have. Thus, multi-band, high valley degeneracy materials with asymmetry in the number of bands, will have the same qualitative influence as asymmetry in $m_{DOS}$.

To fully investigate under which conditions we can observe the increase of the PF when the Femi level is near its intrinsic value in bipolar materials, in Figure 4 and Figure 5 we further explore some important material parameters that can tune and maximize this effect. Understanding the role of simple parameters in maximizing the PF is particularly important in efforts towards the discovery of new materials, when hundreds of possibilities are screened based on simplified parameter descriptors to identify potential high performance materials. [18, 20]

Since what we describe appears in the presence of bipolar transport in narrow bandgap materials, the bandgap is the first parameter we investigate. The role of the bandgap is shown in Figure 4a, where the PF for the asymmetric bandstructure material "A" with bandgap 0.2 eV (as shown in Figure 1d) at $T = 1000$ K is used as the reference case (blue line). Again, the deformation potential for electron-phonon scattering used is $D = 2$ eV for both the CB and VB, and the dielectric constant, which determines ionized impurity scattering screening is $\varepsilon_r = 20$, typical for half-Heuslers. [18] Different bandgaps values, as displayed in the legend of Figure 4a, are simulated for these bands. The dash-dot blue line shows the phonon-limited PF for the material with the smallest bandgap $E_G = 0.2$ eV. As expected, the 'unipolar' PF, i.e. the peak PF in the CB and VB further away from the bandgap, has a negligible dependence on the bandgap. However, the additional spike in the PF observed when $E_F$ is in the bandgap, gradually disappears as the bandgap increases and transport in each band becomes more unipolar. Interestingly, however, bipolar regime PFs similar in magnitude to the unipolar PFs can be achieved even for bandgaps up to $E_G = 0.4$ eV.

In a similar way to the bandgap, the operation temperature determines bipolar transport. We simulate the PF of material "A" with $E_G = 0.2$ eV for three different temperatures $T = 300$ K, 650 K and 1000 K, as shown Figure 4b. In a similar manner to Figure 4a, the blue-dashed line in Figure 4b shows the phonon-limited transport for $T = 1000$ K. As the temperature increases, the bipolar effects become stronger, and the additional spike in the PF becomes higher, reaching its phonon-limited value. Interestingly, the PF increases with temperature across a large region of $\eta_F$ values, because the density of excited minority carriers increases, which in turn corresponds to a lower $N_{imp}$, and higher conductivity. From the temperatures we have simulated, at $T = 650$ K the 'bipolar' PF is already higher compared to the 'unipolar' one.

Complex bandstructure TE materials commonly have CBs and VBs with different degeneracies. In Figure 4c we simulated an example where the VB has double degeneracy and the CB single degeneracy at $T = 1000$ K. The blue and black lines are for the "A" and "S" bandstructure materials as in Figure 1 (as a reference), and "A*" is for the case where the VB is doubly degenerate. Increasing the degeneracy in the VB, moves the $E_F^{intr.}$ closer to the CB, where both the $S$ and $\sigma$ are higher, as in Figure 1b. Thus, the PF becomes even higher, signaling that a highly degeneracy in the minority VB is desired to further increase the PF in majority $n$-type materials.

An additional parameter that influences the asymmetry in transport is the electron-phonon scattering strength, in our case determined by the deformation potentials of the CB and VB. So far we considered only symmetric deformation potentials, but in practice they can differ.[42] The spike in the PF at $\eta_F \sim E_F^{intr.}$ appears when the two conditions, $n = p$ and $n\mu_e = p\mu_h$ are satisfied at different $\eta_F$ values, and the mobilities can provide an additional knob to increase asymmetry through different deformation potentials. In order to focus on the asymmetry of the mobilities rather that the DOS asymmetry, in Figure 4d we compare the PF for the symmetric "S" structure both in terms of bands masses and deformation potentials (black line), with band structures in which the bands are still symmetric, but the deformation potential of the VB is doubled (red line) and tripled (green line). Increasing the deformation potential of the VB moves the $\eta_F$ value for the zero crossing of the Seebeck coefficient closer to the VB (if $\mu_h$ decreases, then $p$ needs to increase to satisfy $n\mu_e = p\mu_h$). On the other hand, $E_F^{intr.}$ still remains at $\eta_F = 0$. As a result, in the $\eta_F \sim E_F^{intr.}$ region, $S$ is nonzero and the high PF spike appears, which further increases as the $D_V$ increases.

Finally, we examine two parameters that influence the degree at which the conductivity reaches phonon-limited values around the charge neutrality point. The first is the amplitude of the deformation potential, which determines the strength of the phonon scattering mechanism in comparison to the strength of the ionized impurity scattering. The blue line in Figure 4e depicts the reference "A" material at $T$ = 1000 K which shows the spike in the PF when the deformation potential is $D$ = 2 eV. The red and green lines show the evolution of the PF when increasing the deformation potential value, to 3 eV and 4 eV. Higher deformation potentials make the phonon-limited conductivity weaker, and reduce the PF spike, as its value at $\eta_F \sim E_F^{intr.}$ is set by its phonon-limited value. Thus, the requirement for the PF spike is that the phonon-limited conductivity value in the $\eta_F \sim E_F^{intr.}$ region is higher compared to the conductivity as determined by the ionized impurity scattering. The PF spike around midgap will be observed more evidently when the material has either a low deformation potential (to increase phonon-limited conductivity), or a low dielectric constant (to reduce ionized impurity limited conductivity).

Similarly, introducing another scattering mechanism such as grain boundary (GB) scattering (as in nanostructured materials) also degrades the spike in the PF. For the PF spike to be evident, and the PF around midgap to be larger compared to the 'unipolar' PF, the strength of all mechanisms together (other than IIS) needs to be weaker compared to the strength of IIS at optimal 'unipolar' conditions. As the IIS weakens with reduced doping, the higher conductivity, limited by the rest of the mechanisms, dominates. Thus, the grain boundary scattering is combined with the phonon scattering mechanism, and degrades the overall conductivity that can be achieved when the IIS strength is diminished around $\eta_F \sim E_F^{intr.}$. In our simplified GB scattering model, by the time the grains have reached sizes as small as 100 nm, the spike in the PF disappears, as shown in Figure 4f.

3.3 *Ultimate PF performance under large asymmetry*

We proceed now by exploring the ultimate PF improvements that can be achieved in the bipolar regime when we boost the electronic asymmetry to a large degree. In Figure 5a we consider how the spike in the PF is improved as the ratio of the deformation potentials between the CB and VB ($D_C/D_V$) changes for different variations in the effective mass of the CB. The effective mass of the VB is kept at $m_0$, its deformation potential at 5 eV, and the dielectric constant at 20. Here the PF gain is evaluated as the ratio between the highest PF achieved within the intrinsic region (spike) and the highest PF achieved if the CB is considered as a unipolar material. The effective mass of the CB is then varied from 0.1 to 1, and its deformation potential of the CB from 1 eV to 10 eV. The calculations are performed at $T = 1000$ K. It is noticeable that the higher the asymmetry, the larger the improvement in the PF, where improvements of more than an order of magnitude are possible for very low deformation potentials and effective masses. Note that the conclusions here do not depend on the type of the band, the band with the lower effective mass and deformation potentials will experience the PF improvement.

Figure 5b similarly explores the impact of variations in the dielectric constant $\varepsilon_r$, together with the deformation potential ratio $D_C/D_V$. This reflects to the relative strength of the IIS processes versus the strength of the phonon scattering processes. The effective masses for the CB and VB are kept constant at 0.2 and 1, respectively, in order to allow a strong further asymmetry from the mass perspective. The deformation potential for the VB is again kept at 5 eV, whereas the deformation potential for the CB is varied from 1 eV to 10 eV and the dielectric constant is varied from 5 to 30. Interestingly, for materials with poor screening, i.e. low dielectric constant, the phonon limited PF value reached in the intrinsic region can be even 40 times higher than what can be estimated by unipolar transport considerations, especially when the deformation potentials are highly asymmetric with lower $D_C$. Note that here we compared the ratio between the spike that appears in the bipolar scheme to the unipolar scheme. If we compute the ratio of the bipolar scheme to the 'combined' unipolar, as computed by eq. 7, we reach even higher improvements because the maximum PF computed in the 'combined' unipolar scheme is lower than the one computed by unipolar considerations alone, as shown in Figures 2c and 2f.

In Figure 5c we show an example for the PF of an *n*-type material which more clearly demonstrates this ultra-high PF improvement. The material parameters are chosen to stretch the PF improvement effect: for the CB we use an effective mass of 0.2 $m_0$ and a deformation potential of 2 eV, and for the VB we use an effective mass of 1 $m_0$ and a deformation potential of 4 eV. The dielectric constant used is 10. With the solid blue line we show the bipolar scheme computation, while the dashed-orange lines show the individual band unipolar calculations, which are then combined using equations 7 to get the bipolar $\sigma$ and $S$, and then multiplied to get the PF (dashed-green line). The enormous PF values indicate that this can be a very promising direction to pursue in TE research. The difference in the PF between the two computational schemes clearly shows how much larger PFs can be predicted when the reduction of $N_{imp}$ is accounted properly.

We have shown how the PF can rise to very high values in the bipolar transport region under extreme anisotropy, but this does not directly translate to improvements in the *ZT* figure of merit. In the bipolar regime, it is expected that the electronic part of the thermal conductivity (including the bipolar thermal conductivity) will increase substantially as well. To examine this, for the material simulated in Figure 5c, we also compute the electronic part of the thermal conductivity $\kappa_e$, evaluated as: [35, 38, 44]

$$\kappa_{e\,ij} = \frac{1}{T}\int_E \Xi_{ij}(E)\left(-\frac{\partial f_0}{\partial E}\right)(E - E_F)^2\, dE - \sigma S^2 T \qquad (8)$$

This is shown in Figure 5d. Again the simulations are performed for $T = 1000$ K. The increase in the electrical conductivity results in a spike in $\kappa_e$ as well, at a degree where the $\kappa_e$ can dominate over the lattice thermal conductivity $\kappa_L$. In fact, we know that the Lorenz number, evaluated as $L = \kappa_e/\sigma T$ in bipolar materials can be substantial, exceedingly overpassing the unipolar values. [45] The Lorenz number, for this material example is shown in Figure 5e. In the heavily doped region it approaches the expected single band value of $L = 2.44 \times 10^{-8}$ W$\Omega$/K$^2$, but in the bipolar region the Lorenz number can reach values as high as 6x compared to the unipolar value.

To observe the overall effect on the *ZT* figure of merit, in Figure 5f we plot for this example material the $ZT = \sigma S^2/(\kappa_e + \kappa_L)$. Here, we take $\kappa_L \sim 2$ W/m·K as a typical value for half Heuslers at high temperatures, [46] but also show with the dashed-dot line the case where only the $\kappa_e$ is considered, as a reference. Despite the ultra-high PF does not directly translate to a *ZT* improvements of the same amplitude, a substantial improvement of more than 2x in *ZT* is achieved at the position of the spike, compared to unipolar considerations when combining the contributions of the valence and conduction band using eq. 7 (dotted blue line for unipolar versus the purple lines in Figure 5f). Importantly, the advantage of a high figure of merit achieved through the PF, rather than low thermal conductivity, is not only beneficial for efficient in energy conversion, but also for offering a higher power output. [11] For applications where the TEG is used at the maximum power output operation, the PF is a quantity of higher importance than the efficiency and *ZT*. Note that in the VB of combined unipolar data in Figure 5d (dotted blue line), the peak tends to shift much further into the VB. It is also evident somewhat in Figure 2c, and amplified in this case of large asymmetry in the bandstructures. It is also related to the fact that the combined unipolar approach of eq. (7) does not capture the $N_{imp}$ density correctly. This exemplifies the necessity to use the fully bipolar simulation scheme compared to combining unipolar calculations in the presence of IIS, as they fail to capture the behavior in the bandgap region (spike) and in the bands (position of PF and *ZT* peak).

4. **Identification of real materials with high PFs**

Here we explore the possibility to observe the spike and large PF in realistic, band asymmetric, bipolar materials. From the most common TE materials, lead compounds PbS, PbTe, PbSe, have very similar effective masses in the CB and VB, and high deformation potentials up to 30 eV. [24] $Bi_2Te_3$, one of the most widely employed and studied TE material, has its CB and VB nearly symmetric, with $m_e \sim 0.08\ m_0$ and $m_h \sim 0.07\ m_0$, [47] and relatively high deformation potentials, between 5 and 16 eV. [48] HgCdTe and HgZnTe can be very interesting materials, because they have narrow and tunable bandgaps with Hg/Cd or Hg/Zn ratio, very asymmetric effective masses and low deformation potentials. [24, 49-52] Nevertheless, the presence of other scattering mechanisms, like polar optical phonons (POP) and alloy scattering, with strengths comparable to the ADP, [50] makes the observation of the spike in the PF in the undoped regime difficult. $AgSe_2$ is another narrow bandgap with very asymmetric DOS effective masses, favoring $n$-type transport, [53] but its transport properties indicate a very high deformation potential value ~ 10 eV. [54]

We have identified a few half-Heusler thermoelectric materials, [18, 23] whose bandstructures fulfill the requirements of achieving high PFs in the undoped regime. For these materials, bipolar transport calculations are conducted using a full-band numerical approach. [18, 23] The first material we have identified is ScNiBi. Its small bandgap of $E_G$ ~ 0.19 eV is confirmed by previous first-principles calculations. [39, 55, 56] Its PF for three temperatures, $T$ = 300 K, 600 K, and 900 K, is plotted versus the Fermi level position $\eta_F$ in Figure 6a, where $\eta_F = 0$ is at the midgap. The specific parameters make transport situation favorable for $p$-type with the high PF spike closer to the VB. The holes in the VB have lighter DOS and conductivity effective mass compared to the conduction band masses, see Tables 1 and 2. In the context of real complex bandstructure materials, the effective mass is extracted as an equivalent effective mass that describes attributes of the whole bandstructure at different temperatures with the corresponding values reported in Table 2. [18] A very large PF spike is observed, reaching values close to PF ~ 50 W/mK$^2$, especially for the higher temperatures, while for the lower temperature, 300 K (blue-diamond line), the PF is still impressive at ~ 25 W/mK$^2$. These values are higher compared to the 'unipolar' PF peak for $\eta_F$ values away from the bandgap into the bands.

Since $\eta_F$ has a direct correspondence to the doping, Figure 6b shows the same data as Figure 6a, but now plotted versus the dopant impurity density, with the left hand side data for acceptor impurities and the right hand side for donor impurities. First, we note that although the lightly doped region with the very high PF values appears to be very narrow when plotted versus $\eta_F$, it actually spans several orders of magnitude of doping concentration and is broad enough to be realizable in experiments. A dopant density below $10^{18}$ cm$^{-3}$ appears sufficient to realize the high

PF region. It is also interesting to observe that because of the high asymmetry in the bands, the zero crossing for the Seebeck coefficient and consequently the PF, appears at the donor side, highlighted by dashed lines in Figure 6b (three lines for the three different temperatures), whereas the acceptor impurity side encloses the entire PF spike. Even with $10^{19}$ cm$^{-3}$ of donors the Seebeck coefficient polarity is still *p* type, despite *n* > *p*.

We have identified three more half-Heusler compounds that exhibit this effect. The PF simulation data versus $N_{imp}$ at 900 K for these materials (and the ScNiBi) are shown in Figure 6c; blue diamonds are for ScNiBi, red circles for ScNiSb, purple stars for NbFeSb, and green asterisks for HfNiSn. The NbFeSb and secondly the ScNiBi exhibit PFs of ~ 60 mW/m·K$^2$ and ~ 40 mW/m·K$^2$ or even higher, respectively. It also appears that a range spanning several orders of magnitude of $N_{imp}$ at both the *n*- and *p*-type sides can be used to reach these incredibly high PFs. It is in principle sufficient to have a density of impurities below $10^{18}$ cm$^{-3}$ in all of these material cases to achieve high PFs. Despite the fact that the ultra-high PF is observed for both acceptor and donor impurity sides, the Seebeck polarity is *p*-type for ScNiBi and ScNiSb, and *n*-type for HfNiSn and NbFeSb. Thus, our results show that the half-Heusler family, with its asymmetric CB/VB, is a promising group to detect the presented PF spikes. The realization of such high PFs is possible in terms of doping values, whereas it is also possible due to the low deformation potentials of these materials. [42] We have used *ab initio* extracted deformation potentials in our calculations as shown in Table 1. Well-established procedures for the evaluation of the deformation potentials from first-principles, enabling predictability, are an ongoing effort in the TE community. [42, 57-59]

It is however important to notice that half-Heusler compounds are usually doped with much higher carrier density than what suggested here, typically ≥ $10^{20}$ cm$^{-3}$, [60-63] whereas indications of best doping in the range around $10^{19}$ cm$^{-3}$ came recently for the (Hf,Zr)NiSn system in polycrystalline samples. [5] It is remarkable that an experimental characterization exists for polycrystalline HfNiSn at the low carrier concentration of ~ 6 x $10^{18}$ cm$^{-3}$, in the range of the suggested values. [64] The measured PF value at 900 K is around 3.6 mW/m·K$^2$, lower than the computed value ~ 15 mW/m·K$^2$. Such difference can be due to the presence of extrinsic defects like grain boundaries (GB), which were specifically introduced to reduce the thermal conductivity. In fact, GB can cover the PF spike, as shown in Figure 4d, or even block the minority carriers, [65] which are the key of the observed PF spike. We specify that we used bandgap values form DFT calculations, which can be underestimated. For ScNiSb, experimental reports indicate a small bandgap ~ 0.2 eV. [66] For HfNiSn and NbFeSb, however, the values in the literature and materials databases vary significantly from 0.3 eV even up to > 1.5 eV. [67] In the case where the bandgaps are larger than ~ 0.8 eV, the effects we describe will not be present. But what we propose is something which can be generic for other low bandgap materials.

After the PF calculations, by computing the electronic part of the thermal conductivity and taking the lattice part from experimental measurements in the literature, we evaluate the figure of merit *ZT*, in Figure 6d for *T* = 900 K. We employed 7 W/m·K for ScNiSb, [66] and use it for ScNiBi as well, 6 W/m·K for NbFeSb, [61] and 4 W/m·K for HfNiSn. [68] Very high *ZT* approaching the values of 3 is computed for *n*-type NbFeSb in the low-doped region. It is notable that HfNiSn, one of the first studied half Heusler compounds, [69, 70] has a high ZT in the highly doped *n*-type region, in agreement with experimental results. [71]

There are still a few recent experimental findings that can signify the proposed effect. In transition metal trichalcogenides undoped single crystals, a bandgap reduction under applied high pressure was observed, accompanied by a PF increase. [72] In the particular case of undoped ZnSe$_3$ crystals, an electron conductivity nearly four times the hole conductivity indicates a huge asymmetry in the electronic properties, favorable to *n*-type transport at ambient conditions. At a certain pressure the main polarity switches to *p*-type, and, upon further pressure increase, the bandgap reduces from 0.25 eV to 0.1 eV and the PF increases by 2 orders of magnitude. This trend is qualitatively similar to our description in Figure 4a.

In another recent work, an increase in the Seebeck coefficient has been detected in undoped Kesterite material, $Cu_2ZnSnS_4$, when it passes through a structural order-disorder transition that results in an increase in the DOS asymmetry (increasing the valence band degeneracy) and a small reduction in the bandgap. [73] Since the electrical resistivity does not change, a rise in the Seebeck's coefficient returns a PF increase. Such PF increase could signal the difference in the Fermi level positions tht satisfy $n = p$ and $\sigma_e = \sigma_h$, quantitatively similar to what we describe in Figures 4a and 4c.

## 5. Conclusions

We have demonstrated that narrow bandgap bipolar TE materials with a high degree of asymmetry between the conduction and valence bands, offer the possibility to achieve extremely large power factor values at lightly doped conditions. Full-band calculations in real TE materials predict that this condition can be achieved within a doping density range of several orders of magnitude. We show that this happens because in these low doping regimes the power factor tends to reach its phonon-limited transport value, with the Fermi level residing at the intrinsic level and allowing high majority carrier electrical conductivity including a significant contribution from the minority carriers. The asymmetry in the conduction/valence bands, which can be in terms of effective masses, deformation potentials, etc., shifts the zero crossing of the Seebeck coefficient away from the intrinsic level, and makes the Seebeck coefficient finite when the conductivity is high. We explained that this effect has not been observed so far, because the most studied thermoelectric materials have rather symmetric bands, though few recent experimental observations can be explained in view of the presented effect. We highlighted the conditions upon which new materials featuring this large PF spike can be discovered. These are low bandgap, bipolar materials with high asymmetry in the DOS and mobility, as well as smaller scattering rates for electron scattering with phonons and any other processes combined, compared to the ionized impurity scattering rates. Finally, we have suggested that some half-Heusler materials, namely HfNiSn, NbFeSb, ScNiBi, ScNiSb, demonstrate this effect and can achieved power factors over 50 W/mK$^2$. Although this high power factor does not translate to high *ZT* in all of these materials (other than NbFeSb with *ZT* ~ 3), still, their *ZT* is significant and it is more beneficial if high *ZT* is achieved through high power factor in high temperature energy harvesting applications.

AUTHOR INFORMATION

**Corresponding Author**

*Patrizio.Graziosi@warwick.ac.uk, Patrizio.Graziosi@gmail.com

**Author Contributions**

The manuscript was written through contributions of all authors.

ACKNOWLEDGMENT

This work has received funding from the Marie Skłodowska-Curie Actions under the Grant agreement ID: 788465 (GENESIS - Generic semiclassical transport simulator for new generation thermoelectric materials) and from the European Research Council (ERC) under the European Union's Horizon 2020 Research and Innovation Programme (Grant Agreement No. 678763).

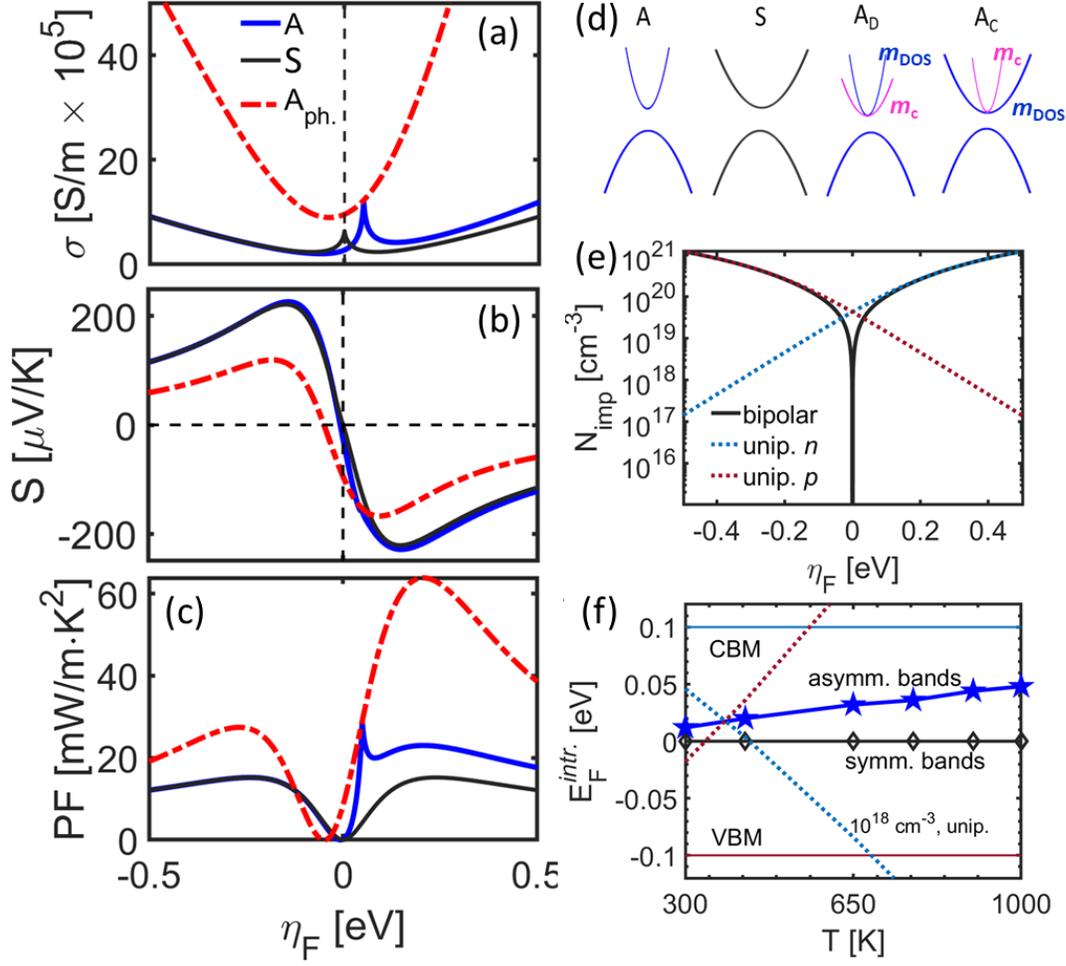

**Figure 1.** (a)-(c) TE coefficients for asymmetric ''A'' (blue) and symmetric ''S'' (black) bands. The dashed-dot red lines indicate the phonon-limited case for bandstructure ''A''. (d) Bands of the asymmetric and symmetric cases. Blue color is for asymmetric bands and black color is for symmetric ones. When the bands are anisotropic (''$A_D$'' and ''$A_C$''), the conductivity effective mass is depicted by magenta lines, whereas the DOS effective mass by the blue lines. (e) Impurity density versus Fermi level position for the symmetric band case, solid black line. The dotted lines show the $N_{imp}$ for the unipolar cases. (f) Temperature dependence of the intrinsic Fermi level $E_F^{intr.}$. CBM refers to the Conduction Band Minimum, VBM to the Valence Band Maximum - blue pentagons are for the asymmetric bandstructure, black diamonds are for the symmetric one. The dotted lines show the position of the Fermi level that gives a carrier density of $10^{18}$ cm$^{-3}$ in unipolar case- blue for *n*-type, red for *p*-type.

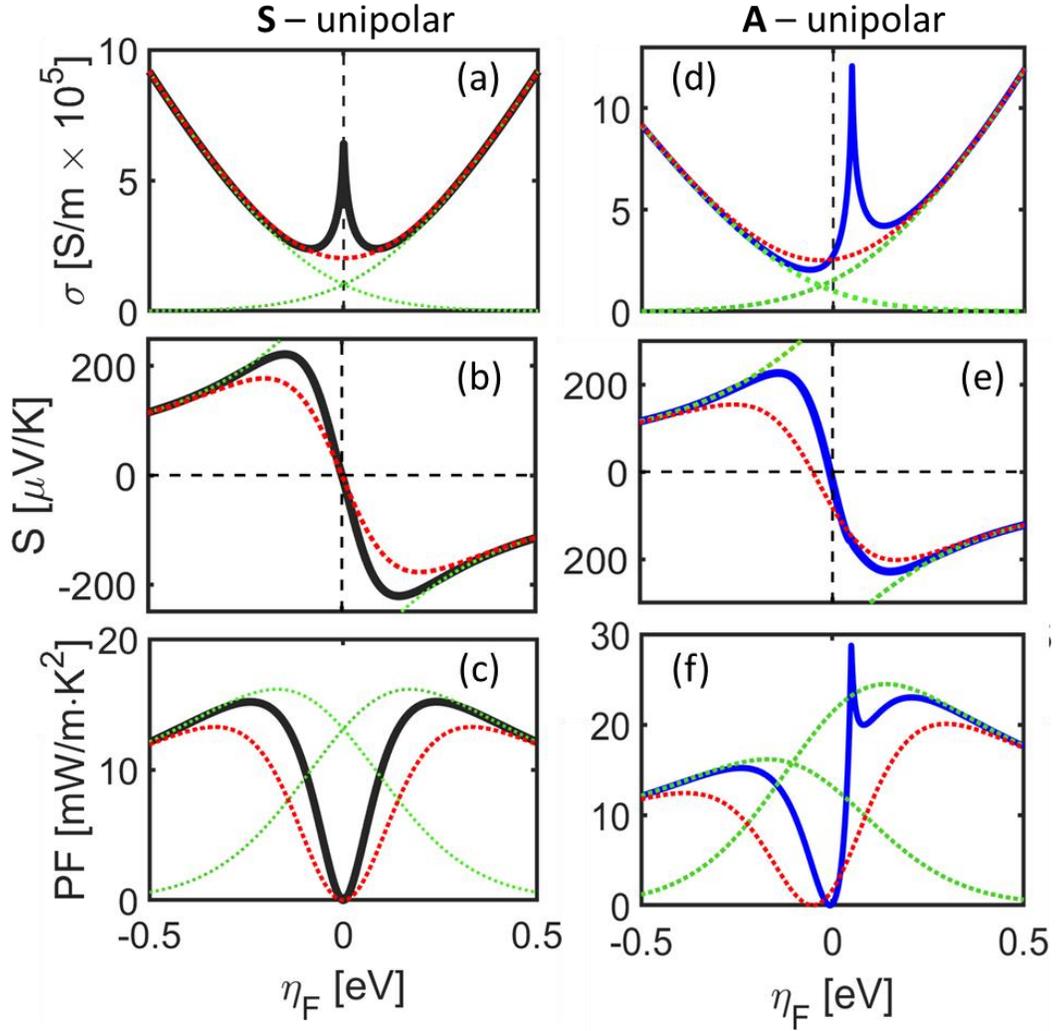

**Figure 2.** TE transport coefficients computed at $T = 1000$ K for the "S" bandstructure, (a) – (c), and for the "A" bandstructure, (d) – (f). Solid lines are for bipolar transport considerations, dotted-green lines are for unipolar transport considerations, which consider separately *n*-type and *p*-type bands, and the dotted-red lines are for the combination of the unipolar calculations. (a) and (d) electrical conductivity. (b) and (e) Seebeck coefficient. (c) and (f) power factor.

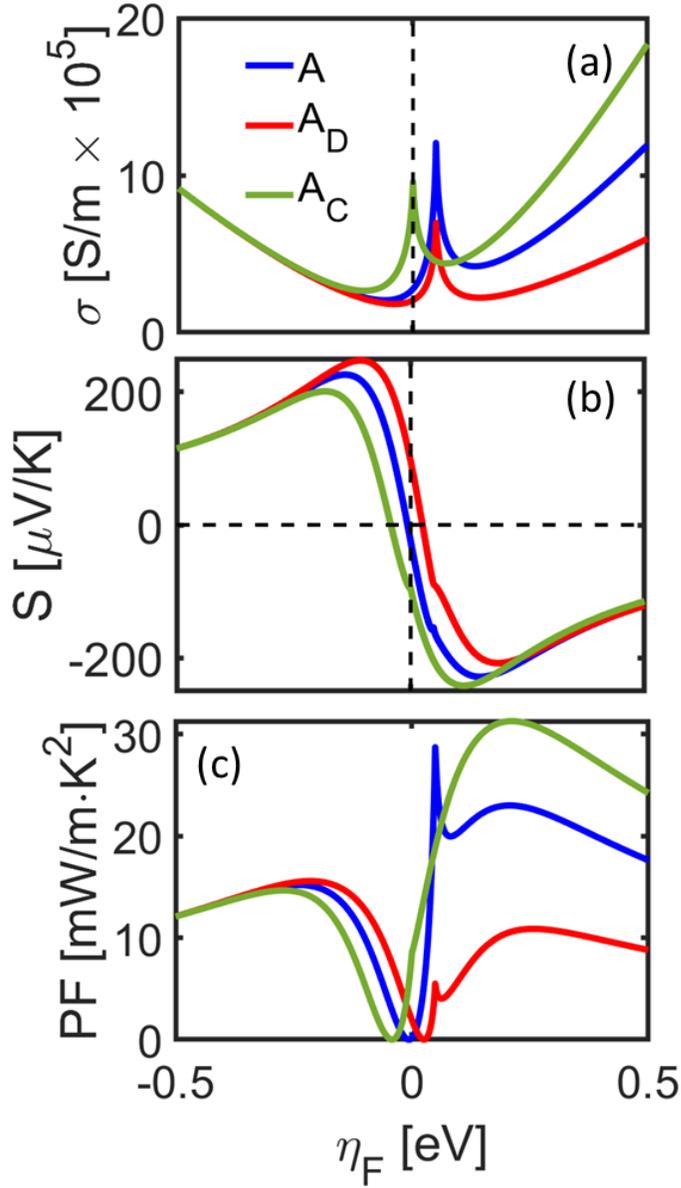

**Figure 3.** TE coefficients (a) electrical conductivity, (b) Seebeck coefficient, and (c) power factor for the case where both the DOS and conductivity effective masses are asymmetric, "A", blue line; compared with the cases where the asymmetry resides only in the DOS effective mass, "$A_D$", red lines; or in the conductivity effective mass, "$A_C$", green lines.

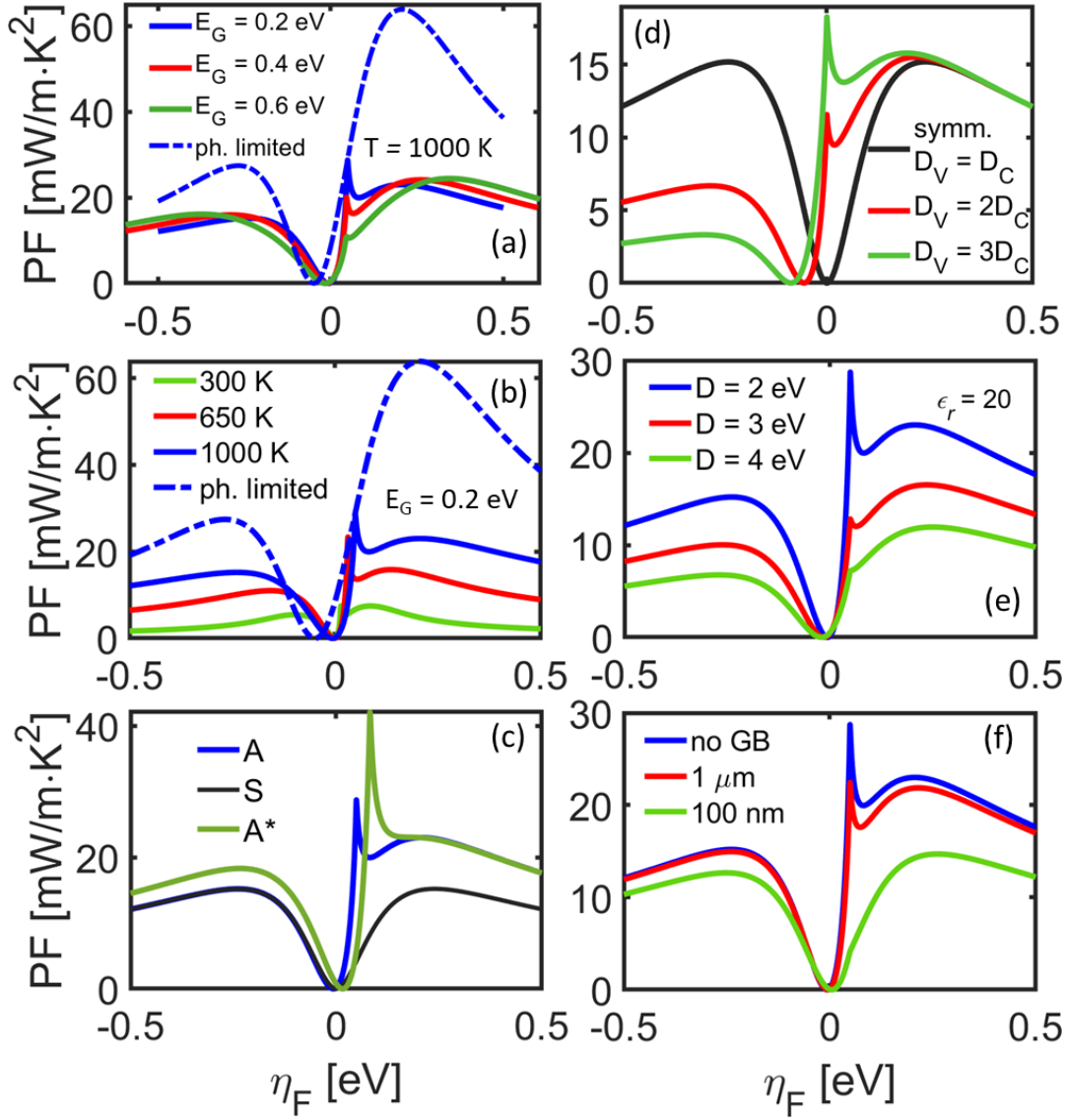

**Figure 4.** (a) TE PF dependence on bandgap at fixed temperature of 1000 K, and (b) on temperature at fixed bandgap of 0.2 eV. (c) The effect of increased DOS asymmetry in the PF where "A*" is for the "A" bandstructure with a double degenerate VB and "S" is for the symmetric bandstructure. (d) PF dependence of the "S" bandstructure material if the deformation potentials are different between VB and CB, $D_C$ = 2 eV, and $D_V$ as in the legend. (e) The PF dependence of the "A" bandstructure material for increasing the deformation potential $D = D_C = D_V$ as displayed in the legend. (f) The PF dependence of the "A" bandstructure when grain boundary scattering with different grain sizes is introduced.

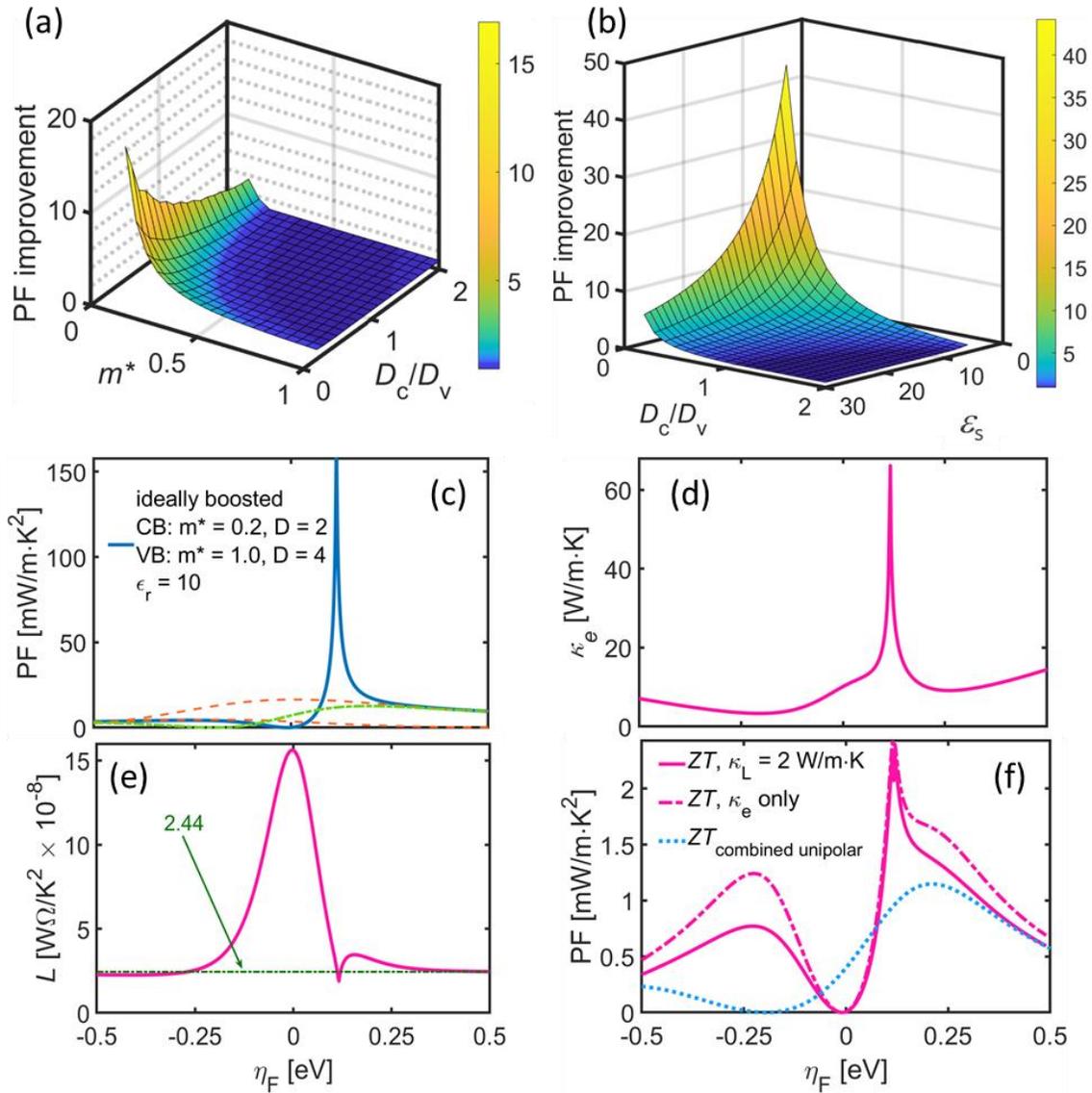

**Figure 5.** Influence of the bandstructures and scattering parameters on the highest PF achievable. The "PF improvement" is evaluated as the ratio between the highest PF achieved with the presented approach (spike) and the PF under unipolar treatment of the separate bands under heavily doped conditions. (a) The influence of the CB effective mass under different deformation potential value ratios $D_c/D_v$. The VB effective mass of is kept at $m_0$, $D_v$ = 5 eV, $\varepsilon_s$ = 20. (b) The effect of changing the dielectric permittivity under different deformation potential value ratio. The effective mass of the CB is fixed at 0.2 and the $D_v$ = 5 eV. (c) PF versus relative position of the Fermi level ($\eta_F$ = 0 is at the midgap) for a material with ideally chosen bandstructure and scattering parameters

as indicated in the legend , to enhance the PF spike; solid line is for the bipolar treatment, dashed-red lines for unipolar calculations, combined in the green-dashed line. (d) Electronic thermal conductivity for the case depicted in (c). (e) and (f) Lorenz number and *ZT*, respectively, for the case depicted in (c). The Lorenz number theoretical reference for degenerate semiconductors value of $2.44 \times 10^{-8}$ W$\Omega$/K$^2$ is highlighted in (e). In (f) the dashed magenta line is the case where only the electronic contribution to the thermal conductivity, $\kappa_e$, is considered, while the dotted blue line is for the *ZT* computed using separate unipolar calculations for CB and VB, and the use of $\kappa_L$. In all these panels $T = 1000$ K is used.

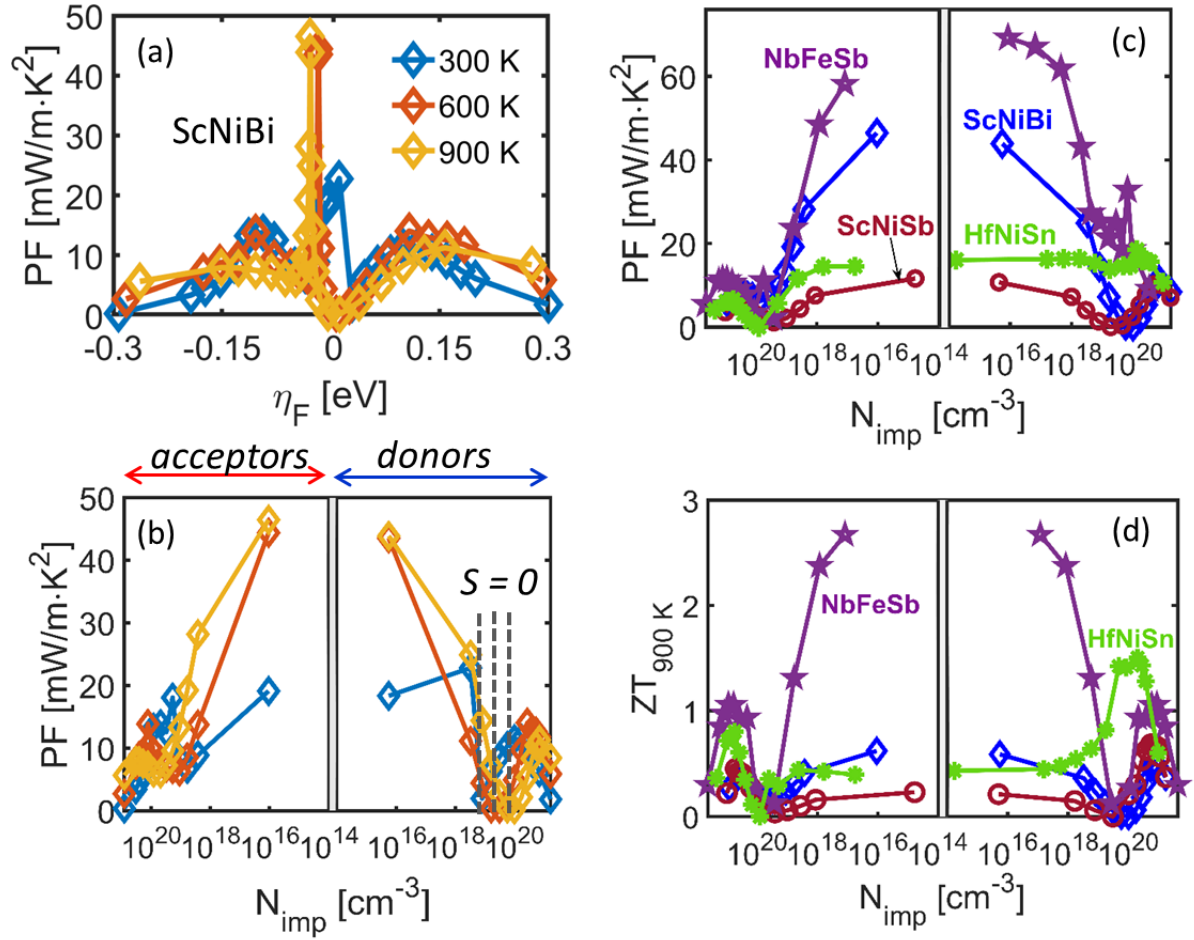

**Figure 6.** Examples of half-Heusler materials where the presented large PF spike in the undoped region can be detected. (a) PF of ScNiBi plotted versus relative Fermi level, and (b) versus the corresponding impurity density. In (a) $\eta_F = 0$ is at the midgap. In (b) the region of $\eta_F$ that corresponds to $N_{imp} < 10^{14}$ cm$^{-3}$ has been removed for graphical clarity. *n*-type data are shown to the right ($\eta_F > 0$) and *p*-type data are shown to the left ($\eta_F < 0$). (c) PF versus impurity density for some half Heusler materials at $T = 900$ K: blue diamonds are for ScNiBi, dark red circles for ScNiSb, purple stars for NbFeSb, and green asterisks for HfNiSn. (d) Figure of merit *ZT* versus impurity density for the half Heusler materials in (c) at $T = 900$ K, using thermal conductivity as detailed in the text.

**Table 1.** The scattering parameters related to the compounds used in Section 4.

| Material | $\rho$ [kg/m$^3$ ×10$^3$] | $u_s$ [m/s ×10$^3$] | ADP [a)] [eV] | ODP [a)] [eV/m ×10$^{10}$] | $\hbar\omega$ [eV] | $\varepsilon_r$ [$\varepsilon_0$] |
|---|---|---|---|---|---|---|
| HfNiSn | 10.3 | 3.39 | 0.1 / 0.7 | 1.8 / 2 | 0.028 | 20.88 |
| NbFeSb | 8.45 | 3.99 | 1 / 0.8 | 1.6 / 2.1 | 0.036 | 22.99 |
| ScNiBi | 8.48 | 3.1 | 0.8 / 0.2 | 2.2 / 1.4 | 0.028 | 29 |
| ScNiSb | 6.6 | 3.9 | 0.7 / 0.4 | 2.3 / 1.8 | 0.032 | 19.66 |

a) electrons / holes

**Table 2.** Extracted bandstructure parameters for the compounds used in Section 4, and DFT computed bandgaps. [18]

| Material | $m_{DOS}$ [a)] | $m_{cond}$ [a)] $T = 300$ K | $m_{cond}$ [a)] $T = 600$ K | $m_{cond}$ [a)] $T = 900$ K | $E_G$ [eV] |
|---|---|---|---|---|---|
| HfNiSn | 1.88 / 4.1 | 0.55 / 0.93 | 0.56 / 0.96 | 0.58 / 1.0 | 0.396 |
| NbFeSb | 0.92 / 5.08 | 0.33 / 1.05 | 0.37 / 1.08 | 0.61 / 1.13 | 0.528 |
| ScNiBi | 3.12 / 1.92 | 0.75 / 0.44 | 0.73 / 0.49 | 0.72 / 0.52 | 0.193 |
| ScNiSb | 3.02 / 2.20 | 0.79 / 0.49 | 0.78 / 0.55 | 0.77 / 0.58 | 0.279 |

a) electrons / holes

TOC graphics

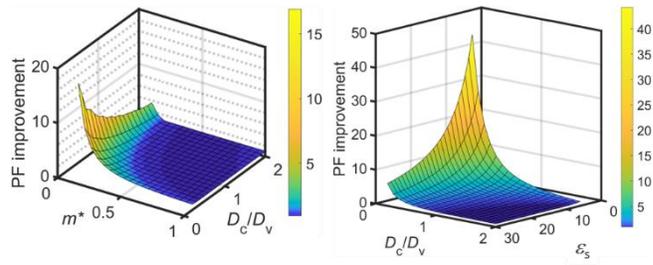